\documentclass[prb,twocolumn,showpacs,aps,superscriptaddress,floatfix]{revtex4-1}
\usepackage{amsmath}
\usepackage{amssymb}
\usepackage{bm}
\usepackage{graphicx}
\usepackage{color}
\usepackage{xcolor}
\usepackage[normalem]{ulem}
\usepackage{hyperref}
\usepackage{float}
\usepackage{verbatim}
\begin{document}
    \title{Nematic superconductivity in topological insulators induced by hexagonal warping}
\author{R.S. Akzyanov}
\affiliation{Dukhov Research Institute of Automatics, Moscow, 127055 Russia}
\affiliation{Moscow Institute of Physics and Technology, Dolgoprudny,
    Moscow Region, 141700 Russia}
\affiliation{Institute for Theoretical and Applied Electrodynamics, Russian
    Academy of Sciences, Moscow, 125412 Russia}

\author{D. A.  Khokhlov}
\affiliation{Dukhov Research Institute of Automatics, Moscow, 127055 Russia}
\affiliation{Moscow Institute of Physics and Technology, Dolgoprudny,
    Moscow Region, 141700 Russia}
\affiliation{National Research University Higher School of Economics, 101000 Moscow, Russia}

\author{A.L. Rakhmanov}
\affiliation{Dukhov Research Institute of Automatics, Moscow, 127055 Russia}
\affiliation{Institute for Theoretical and Applied Electrodynamics, Russian
    Academy of Sciences, Moscow, 125412 Russia}
\affiliation{Moscow Institute of Physics and Technology, Dolgoprudny,
    Moscow Region, 141700 Russia}

    \begin{abstract}
        We study superconducting properties of the bulk states of a doped topological insulator. We obtain that the hexagonal warping stabilizes the nematic spin-triplet superconducting phase with $E_u$ pairing and the direction of the nematic order parameter which opens the full gap is the ground state. This order parameter exhibits non-BCS behavior. The ratio of the order parameter to the critical temperature of $\Delta(0)/T_c$ differs from the BCS ratio. It depends on the chemical potential and the value of the hexagonal warping. We discuss the relevance of the obtained results for the explanation of the experimental observations.
    \end{abstract}


    \maketitle
   
\section{Introduction} 
The non-trivial band structure of topological insulators brings fascinating phenomena such as robust gapless surface states and `topological' electromagnetic response~\cite{Qi2011}. The proximity-induced superconductivity in the topological insulators attracts attention due to possible existence of Majorana fermions~\cite{Fu2008}. Upon doping, the topological insulator becomes a bulk superconductor. Along with s-wave pairing with $A_{1g}$ symmetry, topologically non-trivial pairings with $A_{1u}$, $A_{2u}$, and $E_u$ symmetries become possible~\cite{Fu2010}. The $E_u$ symmetry corresponds to the vector nematic order parameter with the triplet pairing that breaks rotational symmetry. Such a nematic phase supports the Majorana fermions~\cite{Wu2017}, surface Andreev bound states~\cite{Hao2017}, vestigial order~\cite{Hecker2018}, unconventional Higgs modes~\cite{Uematsu2019}. Recently it was shown that the nematic phase may compete with a chiral superconducting phase that breaks time-reversal symmetry~\cite{Huang2018,Kawai2020}.

There is a lot of experimental evidence for the nematic superconductivity with the $E_u$ pairing in doped topological insulators~\cite{Yonezawa2018}. In particular, the data on the Knight shift support the spin-triplet nature of the superconducting order parameter in Cu-doped Bi$_2$Se$_3$~\cite{Matano2016}. Breaking of the rotational symmetry was observed in the heat~\cite{Yonezawa2016}, magnetotransport~\cite{Pan2016}, and STM measurements~\cite{Chen2018,Tao2018}. Recently, it has been demonstrated that the strain dictates the direction of the anisotropy of the second critical field in Sr$_x$Bi$_2$Se$_3$~\cite{Kuntsevich2019} that unambiguously confirms that the ground state in this topological insulator is the nematic superconductivity with $E_u$ symmetry~\cite{Fu2014}. The contact measurements show that the ratio of the superconducting gap to the critical temperature in the doped topological insulator differs from that in the BCS s-wave superconductors~\cite{Kirzhner2012,Tao2018,Sirohi2018}.

From the first sight, the origin of the nematic superconductivity in the doped topological insulators is a mystery. In their seminal work~\cite{Fu2010}, Fu and Berg show that the triplet superconducting order parameter with $A_{1u}$ representation is always favorable in comparison to the nematic order with $E_{u}$ representation and competes with the usual s-wave pairing in topological insulators. Later, it was argued that electron-electron repulsion in Cu$_x$Bi$_2$Se$_3$ favors $A_{1u}$ order~\cite{Brydon2014,Wan2014}, while in Ref.~\cite{Wu2017,Kozii2015}, the authors suggested that a strong Coulomb repulsion between electrons can stabilize the nematic superconductivity. However, this mechanism is doubtful in the case of the topological insulators since the huge dielectric constant in this system implies a weak electron-electron Coulomb interaction, which confirms, in particular, by analysis of the ARPES data~\cite{Chen2013}.

In this letter, we calculate the phase diagram of the doped topological insulator with attractive coupling between charge carriers. Following the approach of Ref.~\cite{Fu2010}, we calculate a superconducting susceptibility of the bulk states to determine the critical temperature of the superconducting phases. Then, we specify our results calculating the free energy. In Ref.~\onlinecite{Fu2010} low-energy expansion of the Hamiltonian has been used taking into account only linear terms to elucidate the main features of the problem. In our calculations, we included the term which is responsible for the hexagonal warping of the Fermi surface. This term is proportional to the third power of the momentum and arises due to the crystal symmetry of the topological insulators~\cite{Fu2009,Liu2010}. The hexagonal warping affects significantly the charge and spin transport in the topological insulators~\cite{Akzyanov2018,Akzyanov2019}. It also is of importance for the characteristics of the nematic superconductivity: the presence of the hexagonal warping gives rise to a full superconducting gap in the spectrum~\cite{Fu2014}. We found that the hexagonal warping stabilizes the superconducting phase with the nematic $E_{u}$ order. If the hexagonal warping is significant, the nematic phase becomes a ground state. This result can explain the observations of nematic $E_{u}$ superconductivity in the experiments with doped topological insulators. We calculate the ratio of the superconducting order parameter at zero temperature to the critical temperature and obtain that, in contrast to the s-wave BCS result, the ratio $\Delta(0)/T_c$ is non-universal and depends on the chemical potential. 
The obtained results can explain the observed non-BCS behavior of superconducting order parameter $\Delta(T)$~\cite{Kirzhner2012,Tao2018,Sirohi2018}. Therefore, the hexagonal warping may be a key for an explanation of the existence of nematic $E_{u}$ superconductivity in real doped topological insulators.


\section{Model Hamiltonian} 
We use the low energy Hamiltonian of the bulk states of topological insulator in the $\mathbf{k}\cdot \mathbf{p}$ model~\cite{Liu2010}. The first-order momentum expansion of this Hamiltonian is 
\begin{eqnarray}
    H_0(\mathbf{k})=m\sigma_x-\mu+v(k_x\sigma_z s_y\!-\!k_y\sigma_z s_x)+v_zk_z \sigma_y,
    \label{Eq::H_0}
\end{eqnarray}
where $s_i$ and $\sigma_i$ $(i=x,y,z)$ are Pauli matrices, $s_i$ acts in the spin space, $\sigma_i$ acts in the orbital space $\mathbf{p}=(P^1,P^2)$, $\mathbf{k}=(k_x,k_y)$ is the momentum, $2m$ is a single electron gap, $\mu$ is the chemical potential, $v$ is the Fermi velocity in the ($\Gamma K$,$\Gamma M$) plane, and $v_z$ is the Fermi velocity along $\Gamma Z$ direction.

Hamiltonian Eq.\eqref{Eq::H_0} is invariant under a continuous rotation in the $(x,y)$ plane. However, crystal structure of a real 3D topological insulator (e.g., Bi$_2$Se$_3$) has only a discrete three fold rotational symmetry and an anisotropic term (that referred to as the hexagonal warping), $H_w(\mathbf{k})=\lambda(k_x^3-3k_x k_y^2)\sigma_z s_z$, appears in the Hamiltonian~\cite{Liu2010}. The total single-electron Hamiltonian is $H(\mathbf{k})=H_0(\mathbf{k})+H_w(\mathbf{k}).$
It obeys an inversion symmetry $PH(\mathbf{k})P=H(-\mathbf{k})$, where the inversion operator is $P=\sigma_x$.

We apply a $U-V$ model to describe the electron-electron interaction. We write down the corresponding term in the Hamiltonian as $H_{int}=-U(n_1^2+n_2^2)-2Vn_1n_2$~\cite{Fu2010}.     
Here $n_i=\sum_{\mathbf{k},s}c^{\dagger}_{i\mathbf{k}s}c_{i\mathbf{k}s}$ and $c^{\dagger}_{i\mathbf{k}s}$ ($c_{i\mathbf{k}s}$) is the creation (annihilation) operator of an electron with momentum ${\mathbf k}$ and spin projection $s$, index $i=1,2$ corresponds to different orbitals $P^i$. The potentials $U$ and $V$ correspond to the intraorbital and interorbital coupling, respectively. We consider the case of attractive interaction, that is, $U,V>0$. Inelastic neutron scattering measurements in Sr$_{0.1}$Bi$_2$Se$_3$ reveal that interorbital electron-phonon coupling can exceed intraorbital one $V>U$~\cite{Wang2019}.

\section{Superconducting order parameters.} In Ref.~\cite{Fu2010} the Hamiltonian $H_0+H_{int}$ (neglecting warping $H_w$) was treated using BCS-like approach. Four possible superconducting pairing symmetries have been classified. The addition of the term $H_w$ does not affect this classification and the results are listed in Table~\ref{my-label}. 

\begin{table}[]
    \centering
    \caption{Possible superconducting pairings taken from Ref.~\cite{Fu2010}}
    \label{my-label}
    \begin{tabular}{|l|l|l|l|l|}
        \hline
        & $\hat{\Delta}_1$ & $\hat{\Delta}_2$ & $\hat{\Delta}_3$ & $\hat{\Delta}_4$
                \\ \hline
        \textrm{Representation} &    $A_{1g}$& $A_{1u}$  &   $A_{2u}$ & $E_u$ \\ \hline
         \textrm{Matrix form}&     1,\,$\sigma_x$ &  $\sigma_ys_z$ &  $\sigma_z$ & $(\sigma_ys_x,\sigma_ys_y)$            \\ \hline
    \end{tabular}
\end{table}

The order parameter $\hat{\Delta}_1$ is even under inversion, while $\hat{\Delta}_2,\hat{\Delta}_3$ and $\hat{\Delta}_4$ are odd under this inversion. The nematic phase, that observed in many experiments, corresponds to the vector order parameter $\hat{\Delta}_4=(\Delta_{4x},\Delta_{4y})=\Delta_4(n_x,n_y)$ in $E_u$ representation. Here vector $\vec{n}=(n_x,n_y)=(\cos \theta,\sin \theta)$ shows a direction of the nematicity. The hexagonal warping affects the gap in the spectrum for the nematic superconductor~\cite{Fu2014}: the warping opens a full gap and this gap is the largest if $\vec{n}=(0,1)$, the nodes exist in the spectrum when $\vec{n}=(1,0)$ even in the case $\lambda\neq 0$.

We use the susceptibilities $\chi_\alpha$ to calculate critical temperature $T_c$ for each possible superconducting phase. The phase with the highest $T_c$ is the ground state. Formally, we can use equations for $T_c$ in a form presented in Ref.~\cite{Fu2010}
\begin{eqnarray}
    \label{Eq::chi_0}
    \hat{\Delta}_1: \det
    \begin{pmatrix}
    U\chi_0(T_c)-1 && U\chi_{01}(T_c)\\
    V\chi_{01}(T_c) && V\chi_1(T_c)-1\\
    \end{pmatrix}
    =0, \\ \nonumber
    \hat{\Delta}_{2,4}: V\chi_{2,4}(T_c)=1, \;\;\;\;\;\;\;\; \hat{\Delta}_3: U\chi_3(T_c)=1,
\end{eqnarray}
where superconducting susceptibilities are ($k_B$=1).
\begin{eqnarray}
    \label{Eq::chi_alpha}
    \chi_{\alpha}(T)&=&\!\!\int\!\! \tanh\!\left(\frac{\xi}{2T}\right)\!d\xi\!\int\!\!\delta(\xi_{\mathbf{k}}-\xi)
    Tr[({\hat m}_{\alpha}Pr_{\mathbf{k}})^2]d^3\mathbf{k}, \\ \nonumber
    \chi_{01}(T)&=&\!\!\int\!\!\tanh\!\left(\frac{\xi}{2T}\right)\!d\xi\!\int\!\!\delta(\xi_{\mathbf{k}}-\xi) Tr[{\hat m}_{0}Pr_{\mathbf{k}}{\hat m}_{x}Pr_{\mathbf{k}}]d^3\mathbf{k},
\end{eqnarray}
the integration is taken over all $\mathbf{k}$. Here $\alpha=0,1,2,3,4x,4y$, notation ${\hat m}_{\alpha}$ represents the matrix structure of the superconducting order parameter in the $\alpha$-phase that given in Table.~\ref{my-label}: ${\hat m}_0=s_0$, ${\hat m}_1=\sigma_x$, ${\hat m}_2=\sigma_ys_z$, ${\hat m}_3=\sigma_z$, ${\hat m}_{4x}=\sigma_ys_x$, ${\hat m}_{4y}=\sigma_ys_y$. Operator $Tr[...]$ is the trace of a matrix and $Pr_{\mathbf{k}}=\sum_{j=1,2}|\phi_{j,\mathbf{k}}\rangle\langle \phi_{j,\mathbf{k}}|$, $\phi_{j,\mathbf{k}}$ is an eigenvector of the Hamiltonian, and $\xi_{\mathbf{k}}$ is the quasiparticle spectrum in the normal state, $(H_0+H_w)\phi_{j,\mathbf{k}}=\xi_{\mathbf{k}}\phi_{j,\mathbf{k}}$. Since $k_z$ enters to integrands only with factor $v_z$, integrals~\eqref{Eq::chi_alpha} are proportional to $1/v_z$ and the phase diagram does not depend on $v_z$.

 We also compute the system free energy at $T=0$ for different order parameter symmetries to verify the results obtained with the help of the superconducting susceptibilities. The phase diagrams near $T=T_c$ and $T=0$ are the same. Note, that values of the order parameters of competing phases are finite at the boundaries of the phase diagram.

\begin{figure*}
    \center{\includegraphics[width=0.95\linewidth]{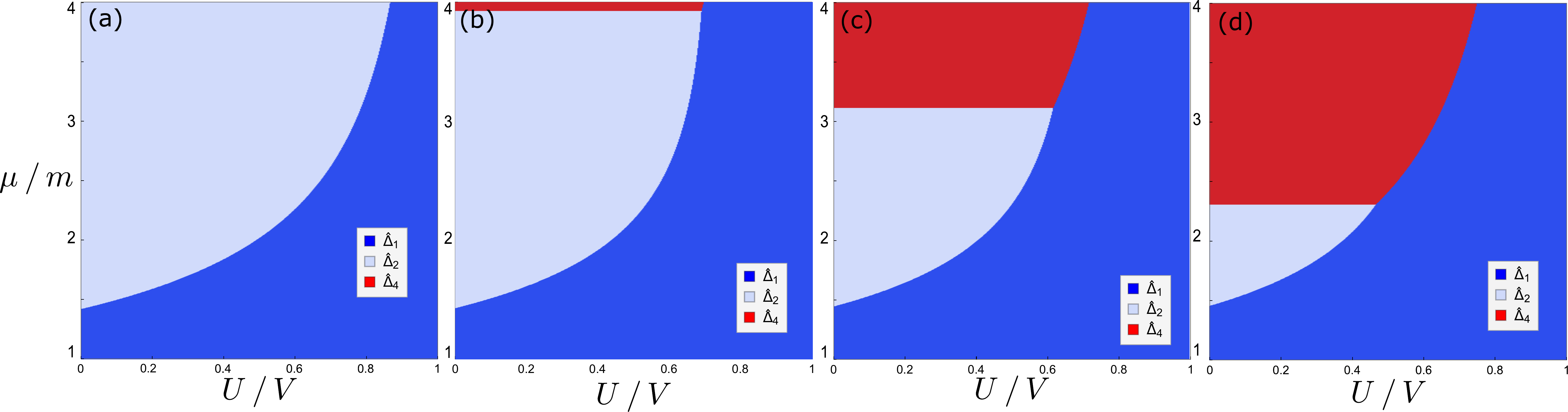}}
    \caption{The phase diagram in the plane $(U/V,\mu/m)$. Blue area corresponds to the ground state with the order parameter $\hat\Delta_1$ (or $A_{1g}$), grey area corresponds to $\hat\Delta_2$ ($A_{2g}$), and red area corresponds to the nematic order parameter $\hat\Delta_4$ ($E_u$). Panels (a), (b), (c), and (d) show phase diagrams for different values of the hexagonal warping, $\lambda m^2/v^3=0,\;0.3, \; 0.5$, and $1$, respectively.}
    \label{Fig::phases_mu_u}
\end{figure*}
\begin{figure*}
\center{\includegraphics[width=0.95\linewidth]{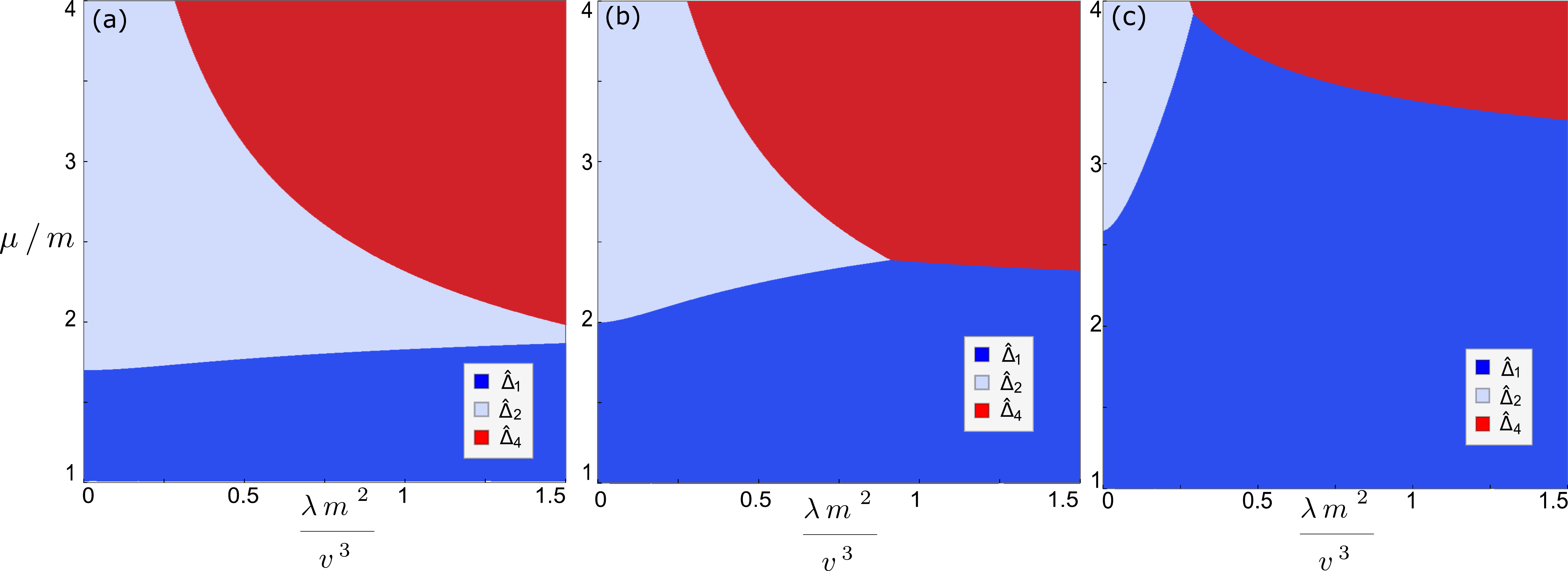}}
\caption{The phase diagram in the plane $(\lambda m^2/v^3,\mu/m)$ for different values of the interaction parameter $U/V$. The notations for the ground state are the same as in Fig.~\ref{Fig::phases_mu_u}. Panels (a), (b), and (c) show the phase diagrams at $U/V=0.3,\;0.5$, and $0.7$ respectively.}
\label{Fig::phases_mu_r}
\end{figure*}

\section{Phase diagram}
A numerical analysis of Eqs.~\eqref{Eq::chi_0} and \eqref{Eq::chi_alpha} with Hamiltonian $H_0+H_w+H_{int}$ shows that  the phase $\hat\Delta_3$ has always lower $T_c$ than $\hat\Delta_1$. Depending on the parameters, the ground state of the system can be $\hat{\Delta}_1$, $\hat{\Delta}_2$ or $\hat{\Delta}_{4}$. In contrast, in the case $\lambda=0$ only $\hat{\Delta}_1$ and $\hat{\Delta}_2$ are candidates for this role~\cite{Fu2010}.  

In the nematic phase $\hat{\Delta}_{4}$, the calculated order parameter and $T_c$ are the same for any direction of nematicity $\vec{n}$. When $\lambda \neq 0$, a full gap in the spectrum opens in the case $\Delta_{4y}$ and the free energy is the smallest for this direction of nematicity. Note, that the difference in the free energy between $\Delta_{4y}$ and $\Delta_{4x}$ is much smaller than a typical difference between the free energies of other phases away from the phase boundaries. It means, in particular, that a small strain can change the relation between $\Delta_{4y}$ and $\Delta_{4x}$ states. We also analyze a possible existence of the chiral phase with order parameter $\Delta_{4y}\pm i\Delta_{4x}$, which spontaneously breaks the time-reversal symmetry. This phase has the same $T_c$ as $\Delta_{4y}$ but higher free energy. However, the chiral phase could be favorable for an extremely small ratio of $v_z/v \ll 1$ and large warping. A similar result was obtained in Ref.~\cite{Chirolli2018} for the model Hamiltonian with a different type of the spectrum non-linearity.

We introduce dimensionless parameters: chemical potential $\mu/m$, hexagonal warping $\lambda m^2/v^3$, and interaction $U/V$. The computed phase diagram of the system is shown in Fig.~\ref{Fig::phases_mu_u} in the plane $(U/V,\mu/m)$ for different $\lambda m^2/v^3$. The singlet pairing $\hat\Delta_1$ is the only ground state if $U>V$. The phase diagram becomes reach when $V>U$. In the absence of the warping, the ground state is either $\hat\Delta_1$ or $\hat\Delta_2$ depending on the chemical potential, Fig.~\ref{Fig::phases_mu_u} (a). In the case of small warping, Fig.~\ref{Fig::phases_mu_u} (b), the nematic phase $\hat\Delta_{4}$ becomes a ground state at large chemical potential and the area with stable nematic $E_u$ paring rapidly increases with the increase of $\lambda m^2/v^3$, Fig.~\ref{Fig::phases_mu_u} (c) and (d). 

The phase diagram in the plane $(\lambda m^2/v^3,\mu/m)$ is shown in Fig.~\ref{Fig::phases_mu_r} for different values $U/V$. As we can see, a moderate ratio between interorbital $V$ and intraorbital $U$ couplings is favorable for the nematic ordering. The increase of the chemical potential benefits both $\hat\Delta_2$ and the nematic $\hat\Delta_{4}$ pairings. The increase of the hexagonal warping makes $\hat\Delta_2$ phase less favorable in comparison to both $\hat\Delta_1$ and nematic pairings. The growth of $\mu/m$ stimulates nematic phase $\Delta_{4y}$ as compared to the pairing $\hat\Delta_1$, especially, at small values of intraorbital interaction $U$. 

The nematic pairing $\hat{\Delta}_{4}: (c_{k1\uparrow}c_{-k2\uparrow} + c_{k1\downarrow}c_{-k2\downarrow})$~\cite{Fu2014}, couples electrons in different orbitals with the same spin. If we neglect the hexagonal warping, the numbers of electrons with opposite spin projection are equal at each orbital. The hexagonal warping $H_w=\lambda(k) s_z\sigma_z$ shifts this spin projection balance: the moving electrons obeying given orbital become polarized, while the total polarization of electron liquid remains zero. Thus, the hexagonal warping favors the nematic superconductivity since it creates a favorable spin-orbit configuration.

\section{Mean field calculations for $E_u$ pairing.} 
\begin{figure}
    \centering
    \includegraphics [width=8.5cm]{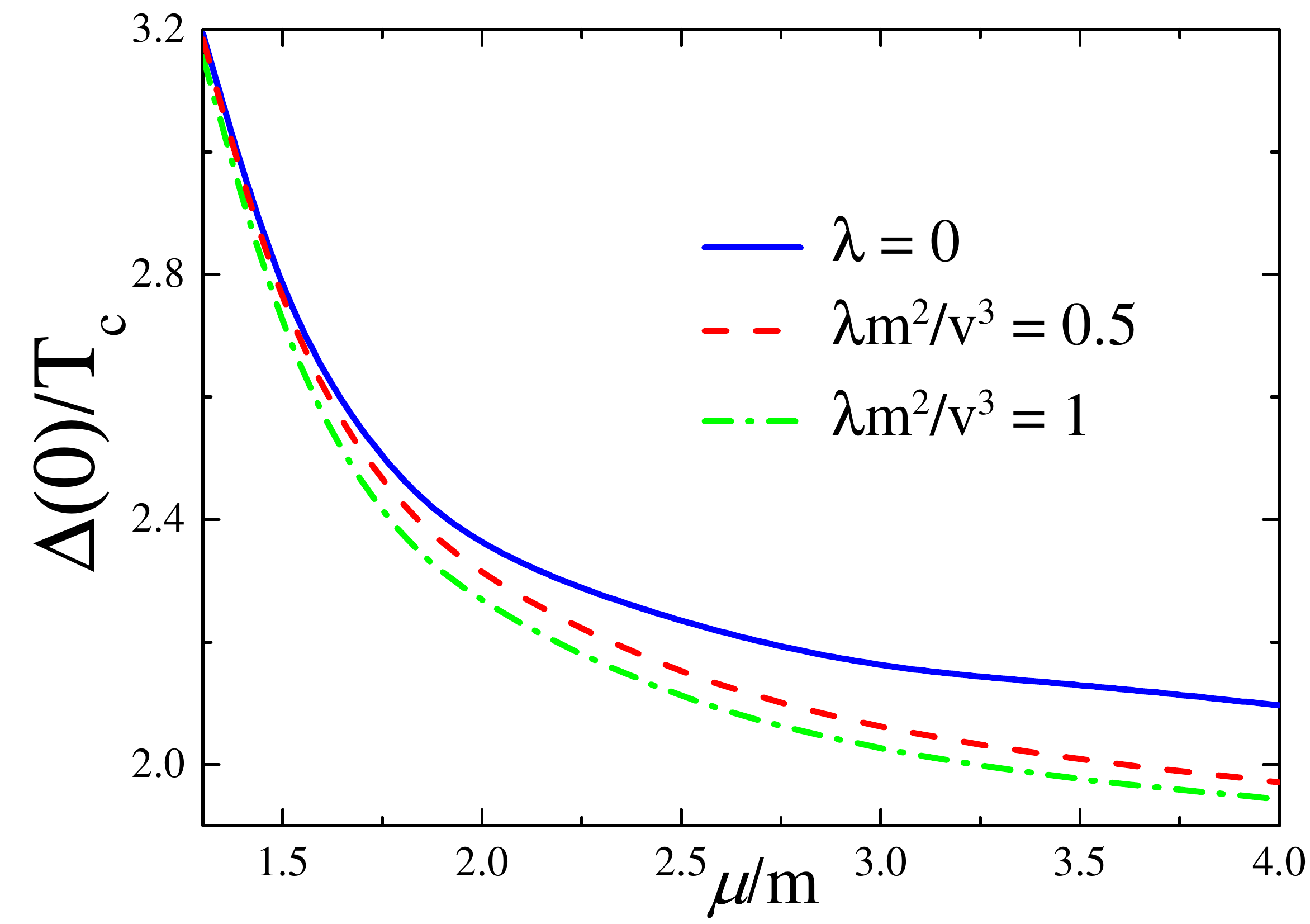}
    \caption{Ratio of the order parameter at zero temperature and the critical temperature, $\Delta(0)/T_c$, as a function of the chemical potential for different values of the hexagonal warping. Blue line corresponds to $\lambda=0$, red dashed line to $\lambda m^2/v^3=0.5$, and green dot dashed line to $\lambda m^2/v^3=1$.}
    \label{dtcu}
    \end{figure}

\begin{figure}
        \centering
       \includegraphics [width=8.5cm]{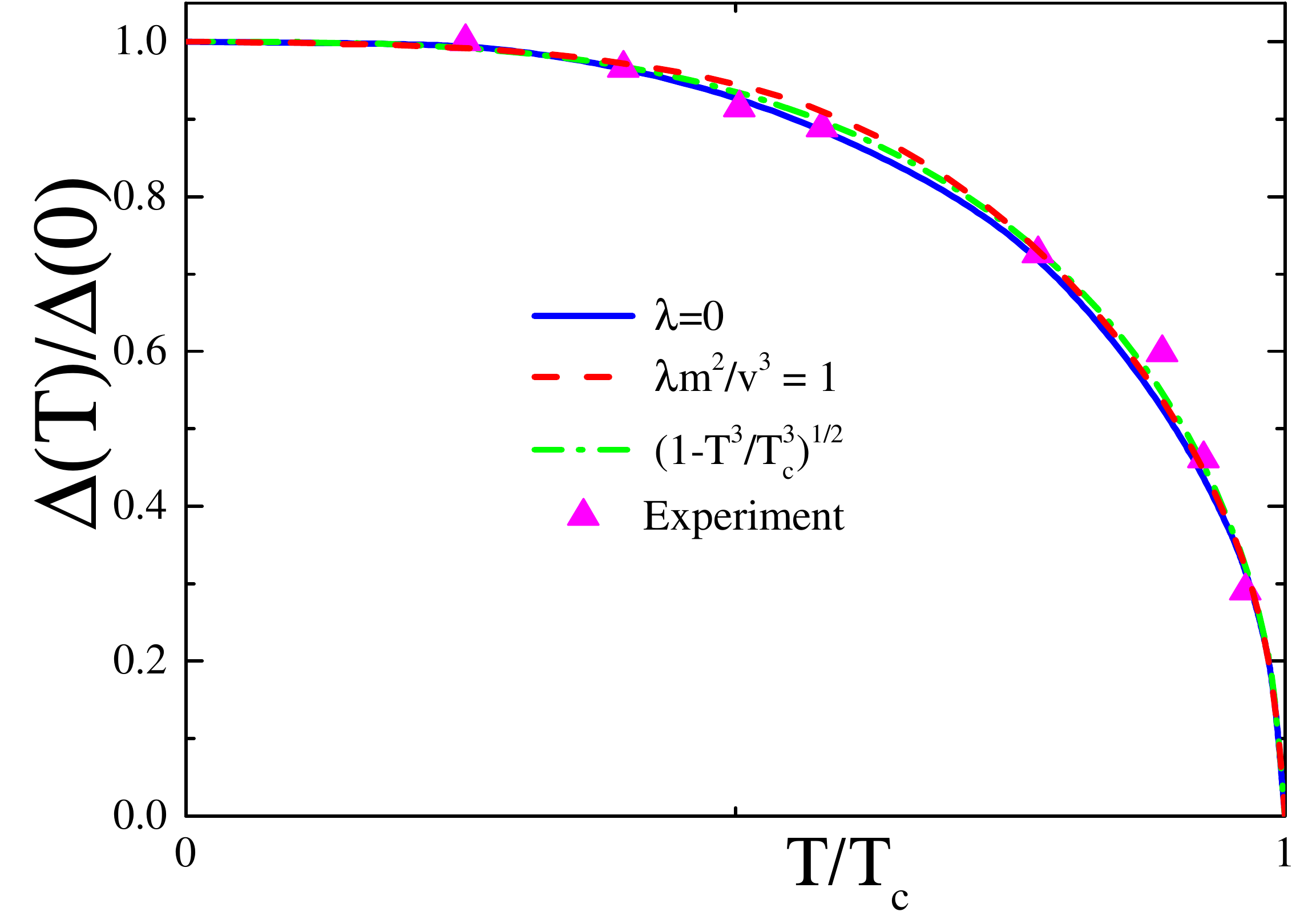}
        \caption{Dependence of the order parameter $\Delta$ on temperature for different values of the hexagonal warping. The blue line corresponds to zero warping, red line to $\lambda m^2/v^3=1$, both curves are plotted at $\mu/m=2$. Green dot-dashed line is a fit $\Delta(T)/\Delta(0)=\sqrt{1-\left(T/T_c\right)^3}$. Purple triangles correspond to the experimental data extracted from Ref.~\cite{Sirohi2018}.}
        \label{DeltaT}
\end{figure}

Here we calculate the absolute value $\Delta$ of the nematic order parameter $\hat\Delta_{4}=\Delta(0,1)$. We choose the Nambu basis as $\Psi_{\mathbf{k}}=(\phi_{\mathbf{k}},-is_y\phi^{\dagger}_{-\mathbf{k}})^t$,
where $\phi_{\mathbf{k}}=(\phi_{\uparrow,1,\mathbf{k}},\phi_{\downarrow,1,\mathbf{k}},\phi_{\uparrow,-1,\mathbf{k}},\phi_{\downarrow,-1,\mathbf{k}})^t$ and the superscript $t$ means transposition. In this basis the mean-field Hamiltonian of the topological insulator with the nematic superconducting order can be written as
\begin{eqnarray}\label{hbdg}
H_{BdG}(\mathbf{k})=(H_0+H_w)\tau_z+\Delta\sigma_ys_y\tau_x,
\end{eqnarray}
where $\tau_i$ is the Pauli matrix in the electron-hole space. Taking in mind that in terms of the creation-annihilation operators the nematic order corresponds to $c_{1\sigma}c_{2\sigma}$ pairing~\cite{Fu2010}, we can write the mean-field free energy in the form 
\begin{equation}
\Omega=\frac{2\Delta^2}{V}-2T\sum\limits_i\int \frac{d^3\,k}{(2\pi)^3}\ln\left\{1+\exp{\left[-\frac{\epsilon_i({\mathbf k})}{T}\right]}\right\},
\end{equation}
where $\epsilon_i({\mathbf k})$ is the $i$-th eigenvalue of $H_{BdG}$. We compute $\Delta(T)$ by minimizing $\Omega$. This calculations reveal that the ratio $\Delta(0)/T_c$ is not a constant in contrast to the s-wave BCS superconductivity and depends on the chemical potential and the warping, see Fig.~\ref{dtcu}. The value $\Delta(0)/T_c$ decreases with $\mu$ and $\lambda$ and may be considerably larger than the BCS value $1.76$. In Fig.~\ref{DeltaT} we show that the dependence of $\Delta(T)/\Delta(0)$ on temperature can be approximated as $\Delta(T)/\Delta(0)\approx \sqrt{1-\left(T/T_c\right)^3}$ for all range of temperatures and is almost independent of the value of hexagonal warping. In Ref.~\cite{Sirohi2018} the dependence of the gap in Nb$_x$Bi$_2$Se$_3$ on temperature has been measured. The results are shown in Fig.~\ref{DeltaT} by pink triangles. As we can see from this figure, the theory fits the experiment well. 

\section {Discussion.} We show that the existence of the hexagonal warping can explain the experimental observations of the $E_u$ superconducting pairing in the doped topological insulators if the dimensionless parameters $\lambda m^2/v^3$ and $\mu/m$ are not small. In the undoped Bi$_2$Se$_3$, the chemical potential lies near the band edge $\mu/m=1.3$ and the strength of the warping is estimated as $\lambda m^2/v^3=0.11$~\cite{Liu2010}. These values are too small for the existence of the nematic phase. However, the doping by Cu or Sb can significantly increase the chemical potential and the value $\mu/m = 2$ or larger looks realistic~\cite{Lahoud2013,Neupane2016}. The Fermi velocity also significantly affects the warping parameter $\lambda m^2/v^3$. The reported Fermi velocities for the surface states in the topological insulators lie in a wide range from $v = 5\cdot10^7$~cm/s in Ref.~\cite{Zhang2009} to $v =10^7$~cm/s in Ref.~\cite{Veldhorst2012} and even down to $v =3 \cdot 10^5$~cm/s in Ref.~\cite{Wolos2012}. Thus, the necessary large value of the effective hexagonal warping is realistic. For example, if the Fermi velocity for the bulk states $v$ decreases by half in comparison with DFT calculations for the undoped Bi$_2$Se$_3$, then, the warping parameter increases by eight, $\lambda m^2/v^3 \approx 0.9$, and the nematic phase become favorable. It is hard to estimate the ratio of the intraorbital to interorbital attraction $U/V$. In Ref.~\cite{Wan2014} the electron-phonon couplings have been calculated for Cu$_{0.16}$Bi$_2$Se$_3$ using DFT approach under an assumption of a weak effect of the Cu doping on the structural properties. It has been obtained that the triplet pairing $A_{2g}$ has lower free energy than $A_{1g}$. In our terms, it means that $V>U$, which is necessary for the nematic $E_u$ pairing.  However, the X-ray experiments show that even small doping has a considerable effect on the structural properties of the topological insulators~\cite{Kuntsevich2018,Kuntsevich2019}, so the values $v$ and $\lambda$ can be also affected by doping.

The gap in the energy spectrum $2\bar\Delta$ relates to the order parameter as $\bar\Delta = \Delta \lambda k_F^3/\mu$ when $\Delta\ll v/k_F$, where $k_F$ is the Fermi momentum in the normal state in $\Gamma K$ direction. In Ref.~\onlinecite{Tao2018} the superconducting gap in Cu$_x$Bi$_2$Se$_3$ was estimated as $\bar\Delta \approx 0.256$~meV in sample with $T_c\sim 3$K. If we take typical values for the nematic superconductivity $\lambda m^2/v^3 =0.5$ and $\mu/m=3$ (see Fig.~\ref{Fig::phases_mu_r}), then, we get $\bar\Delta \approx 0.35\Delta$. Using these parameters we get an estimate $\Delta(0)/T_c\approx 2.4$ for Cu$_x$Bi$_2$Se$_3$ samples, which is in a good agreement with the result shown in Fig.~\ref{dtcu}.

The Anderson theorem is valid in the case of considered topological superconductivity~\cite{Anderson2020}, and weak disorder does not affect significantly the obtained results. The influence of fluctuations on nematic superconductivity has been studied in Ref.~\onlinecite{Wu2017} using a simplified parabolic Hamiltonian. The fluctuations generate attraction in both the s-wave and nematic channels, while the fluctuations in $A_{1u}$ and $A_{2u}$ channels stimulate the s-wave pairing. Thus, we expect that fluctuations increase the region of the nematic and s-wave phases in the phase diagram.

\section{Conclusions}
We suggest a possible mechanism of the ground-state nematic superconductivity with the spin-triplet $E_u$ pairing which observed in doped topological insulators. We show that the hexagonal warping is an essential feature for the realization of the nematic superconductivity. The nematic superconductivity has a non-BCS behavior of the order parameter. In particular, the ratio of the order parameter at $T=0$ to the critical temperature is non-universal and depends on the chemical potential and warping. 

\section{Acknowledgements} 
ALR acknowledges the support by JSPS-RFBR Grant No. 19-52-50015 and RFBR Grant No. 19-02-00421. RSA and DAK were supported RSA by the Russian Scientific Foundation under Grant No 20-72-00030 and the Foundation for the Advancement of Theoretical Physics and Mathematics “BASIS”.

\bibliography{bib}
\end{document}